\documentclass[conference]{IEEEtran}
\IEEEoverridecommandlockouts
\usepackage{cite}
\usepackage{amsmath,amssymb,amsfonts}
\usepackage{algorithmic}
\usepackage{graphicx}
\usepackage{textcomp}
\usepackage{xcolor}
\usepackage{svg}

\def\BibTeX{{\rm B\kern-.05em{\sc i\kern-.025em b}\kern-.08em
    T\kern-.1667em\lower.7ex\hbox{E}\kern-.125emX}}
\begin{document}

\title{Ground vehicle odometry using a non-intrusive inertial speed sensor\\
}

% \author{\IEEEauthorblockN{Het Shah}
% \IEEEauthorblockA{
% \textit{Mechanical Engg. Dept.} \\
% \textit{Indian Institute of Technology}\\
% %  \texetit{Indian Institute of Technology, Kharagpur}\\
% Kharagpur, india \\
% hetshah369@iitkgp.ac.in}
% \and
% \IEEEauthorblockN{Siddhant Haldar}
% \IEEEauthorblockA{
% \textit{Dept. of Electrical Engg.} \\
% \textit{Indian Institute of Technology} \\
% % \textit{Indian Institute of Technology, Kharagpur}\\
% Kharagpur, india \\
% siddanthaldar14@iitkgp.ac.in}
% \and
% \IEEEauthorblockN{Rohit Ner}
% \IEEEauthorblockA{\textit{Dept. of Mathematics} \\
% \textit{Indian Institute of Technology} \\
% % \textit{Indian Institute of Technology, Kharagpur}\\
% Kharagpur, india \\
% rohitner@iitkgp.ac.in}
% \and
% \IEEEauthorblockN{Siddharth Jha}
% \IEEEauthorblockA{
% \textit{Dept. of Electrical Engg.} \\
% \textit{Indian Institute of Technology} \\
% %\textit{Indian Institute of Technology, Kharagpur}\\
% Kharagpur, india \\
% thesidjway@iitkgp.ac.in}
% \and
% \IEEEauthorblockN{Dr. Debashish Chakravarty}
% \IEEEauthorblockA{
% \textit{Dept. of Mining Engg.} \\
% \textit{Indian Institute of Technology} \\
% %\textit{Indian Institute of Technology, Kharagpur}\\
% Kharagpur, india \\
% dc@mining.iitkgp.ac.in}

% }

\author{\IEEEauthorblockN{\IEEEauthorrefmark{1}Het Shah\IEEEauthorrefmark{2},
\IEEEauthorrefmark{1}Siddhant Haldar\IEEEauthorrefmark{3},
\IEEEauthorrefmark{1}Rohit Ner\IEEEauthorrefmark{4}, 
\IEEEauthorrefmark{1}Siddharth Jha\IEEEauthorrefmark{5},
Debashish Chakravarty\IEEEauthorrefmark{6}}
\IEEEauthorblockA{\IEEEauthorrefmark{2}Department of Mechanical Engineering,\IEEEauthorrefmark{3}\IEEEauthorrefmark{5}Department of Electrical Engineering,\\\IEEEauthorrefmark{4}Department of Mathematics,\IEEEauthorrefmark{6}Department of Mining Engineering}
Indian Institute of Technology, Kharagpur, India \\
\tt \{\IEEEauthorrefmark{2}hetshah369, \IEEEauthorrefmark{3}siddhanthaldar14, \IEEEauthorrefmark{4}rohitner, \IEEEauthorrefmark{5}thesidjway\}@iitkgp.ac.in\\
\IEEEauthorrefmark{6}dc@mining.iitkgp.ac.in
\thanks{\IEEEauthorrefmark{1} denotes equal contribution.}
}

\maketitle

\begin{abstract}
This paper describes the design and development of a non-intrusive inertial speed sensor that can be reliably used to replace a conventional optical or hall effect-based speedometer on any kind of ground vehicle. The design allows for simple assembly-disassembly from tyre rims. The sensor design and data flow are explained. Algorithms and filters for pre-processing and processing the data are detailed. Comparison with a real optical encoder proves the accuracy of the proposed sensor. Finally, it is shown that factor graph-based localization is possible with the developed sensor.
\end{abstract}


\section{Introduction}
Localization of ground robots and cars has been a problem which has been studied for a while. Since the research on self-driving cars picked up the pace, the localization problem has been studied even more rigorously. The earliest successful autonomous cars like Stanley \cite{thrun2006stanley} had algorithms to fuse data from GPS, GPS compass, IMU (Inertial Measurement Unit), and the wheel encoders. The same principle is followed today in most autonomous vehicles. As far as the optical wheel encoders are concerned, they are long known as reliable sources of accurate and precise velocity information, and are usually mounted around the shaft of the wheel. All modern cars are fit with a speedometer, which usually works using data from Hall-effect sensors, which read magnetic fields from magnets mounted on the car's rotating driveshaft to estimate speed \cite{Woodford09speedo}. Accurate and precise information about a car's speed is readily available using these sensors. But a major setback for both these sensors is that they are rigidly mounted during a car's construction, and thus are not easily replaceable. In other words, one cannot easily remove and replace a car's speed measurement devices without significant disassembling. 

The motivation behind the development of a non-intrusive inertial speed sensor was a drive-by-wire system, mounted on a Mahindra E2O electric car, which our team at IIT Kharagpur is in the process of converting into a self-driving car. The system outputs data of all the sensors present in the car, including the data of the speedometer, but the speedometer data has a least count of 1 kmph (~0.28m/s) and a range from 1-90 kmph. Although such precision is not an issue for manual transmission at high speeds, autonomous driving (which is usually done at low speeds for obvious reasons, averaging 2m/s), requires more precise speed data. This paper describes the design and development of such a non-intrusive sensor that can be mounted on the wheel rim of any car, and provide good quality speed estimates at good precision. 

The idea of such a non-intrusive sensor is not entirely new. A similar sensor in which an accelerometer is mounted on the rim of a wheelchair wheel is presented in \cite{coulter2011development}. \cite{sonenblum2012validation} uses a physical activity monitor mounted on a wheelchair to measure its movement. \cite{Gersdorf2013AKF} is a novel work that uses an Extended Kalman Filter for a wheel-mounted inertial measurement unit (IMU) using two accelerometers and a single gyroscope as a substitute for optical/magnetic encoders. However, usage of such a sensor for localizing a robotic car is novel to our knowledge. This paper describes the design and development of the sensor, evaluation of its performance with comparison to a standard optical wheel encoder and its usage for localization of the ground vehicle.
%https://www.ncbi.nlm.nih.gov/pubmed/22698978
%https://www.nature.com/articles/sc2010126

Factor graph optimization \cite{loeliger2004introduction} is getting increasingly popular for localization problems as an alternative to filter-based approaches. Factor graphs are recently being used to solve large-scale vision and LASER-based localization \cite{forsterimu}, but are flexible enough to solve the traditional wheel encoder + GPS-based localization problem as described in this paper.

%http://people.binf.ku.dk/~thamelry/MLSB08/hal.pdf

This paper is divided into six sections. Section \ref{sec:nonintrusive} describes the sensor functionality and design principles, along with the data flow. Section \ref{sec:dataproc} describes the bias estimation and removal algorithm used, and the state estimation using two techniques namely Kalman Filtering and Recursive Least Squares. Section \ref{sec:localize} describes the factor graph method for GPS + Inertial sensor fusion for localization of the car. In section \ref{sec:experiment}, we demonstrate the performance of the sensor in real-world tests and compare it with an optical encoder's performance as a benchmark. Finally, we conclude this paper in section \ref{sec:conclusion}.

%https://ac.els-cdn.com/S1350453312001294/1-s2.0-S1350453312001294-main.pdf?_tid=5383736f-b37d-4841-9cd0-43723552d5f7&acdnat=1536861731_cf02f4dd838f7cff1ee9c27746c42425
%https://www.nature.com/articles/sc2010126.pdf

\section{Non-Intrusive Inertial Speed Sensor}
\label{sec:nonintrusive}

\begin{figure}[h]
\centering
\noindent \includegraphics[width=0.95\linewidth]{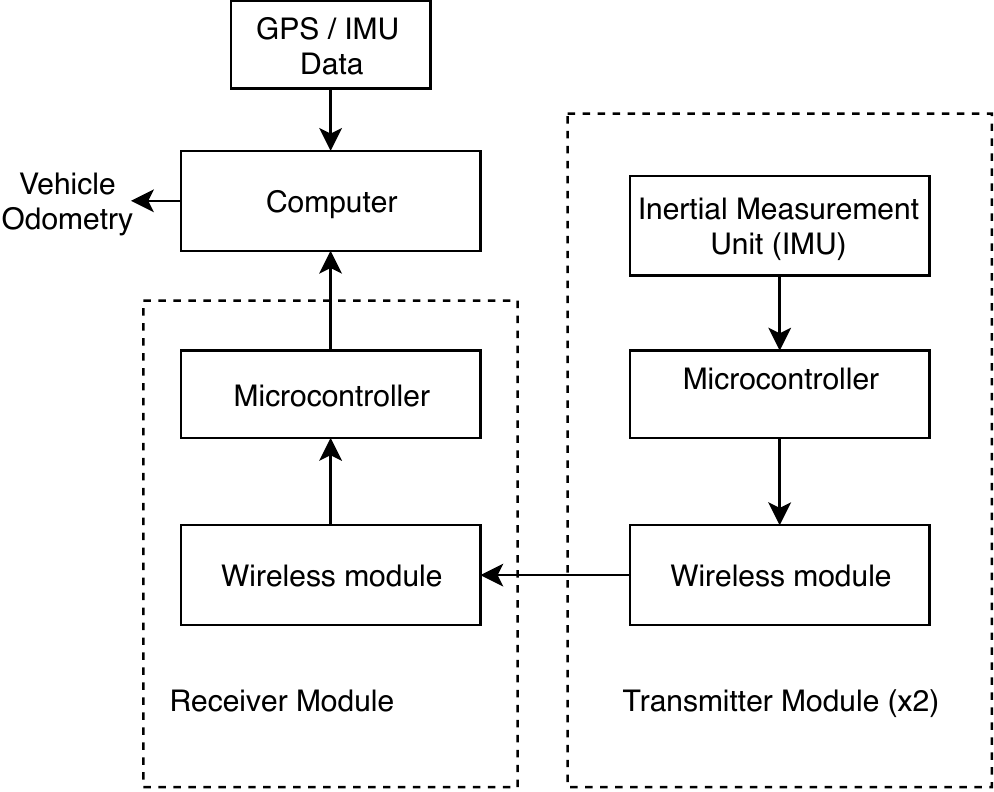}
\caption{Data flow in the experimental setup}
\label{dataflow}
\end{figure}

To acquire wheel speed measurement data from the mobile robot, a non-intrusive inertial speed sensor was developed. An Inertial Measurement Unit(IMU), comprising of a 3-axis gyroscope (to measure angular velocity), a 3-axis accelerometer (to measure linear acceleration) and a 3-axis magnetometer (to measure magnetic field strengths) is used as the main measurement device in this sensor. The sensor can be mounted on the rim of a robotic car's rear wheels as a replacement for intrusive wheel encoders, although its scope is not limited to the same. The sensor is designed to be power-efficient and small in size. Its modular design allows it to be mounted on most kinds of cars and other land vehicles with minimal effort and changes to the code base. The wireless nature of the sensor allows its data to be processed on any computer in the vicinity, mounted either on the robot or off it. Figure \ref{dataflow} shows the main components of the experimental mobile robot car setup, consisting of the computer, the wheel-mounted inertial speed sensors, microcontrollers, and Inertial Navigation Sensor(INS - IMU + GPS).

% \begin{figure}[b]
% \centerline{\includegraphics{Images/dataflow.pdf}}
% \caption{Data flow in the experimental setup}
% \label{fig}
% \end{figure}

\begin{figure}[h]
\centering
\noindent \includegraphics[width=0.7\linewidth]{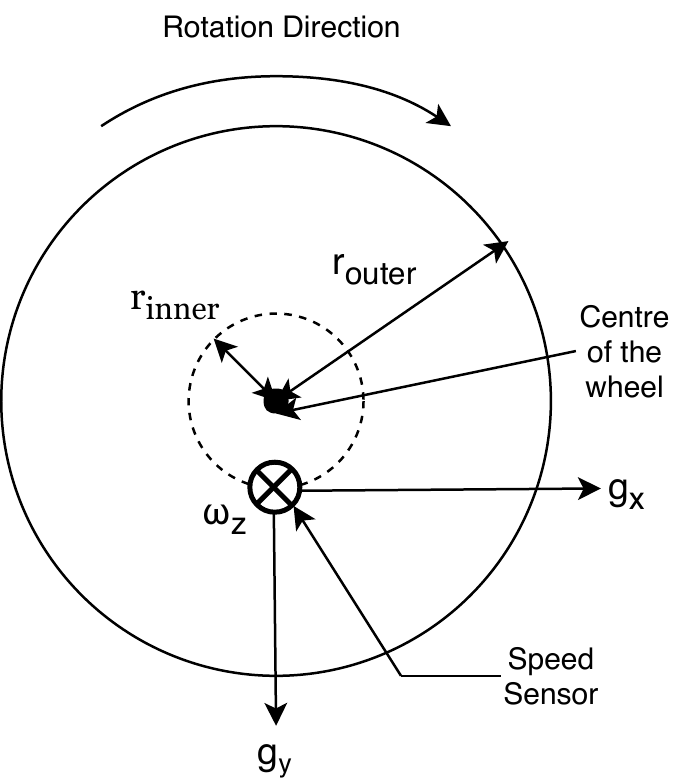}
\caption{Sensor coordinate system and placement}
\label{sensorcoord}
\end{figure}

To maintain the modularity of the sensor, it is powered onboard by a portable rechargeable cell. A charging circuit for the cell is also present on the sensor PCB. The coordinate system and placement of the sensor on the vehicle's wheel are shown in Figure \ref{sensorcoord}, showing that the sensor can be placed anywhere on the rim. A detailed description of the circuit components is given in section \ref{sec:experiment}.

\section{Data Processing}
\label{sec:dataproc}
\subsection{Bias Estimate Initialization}\label{AA}
Typical Micro Electro Mechanical Systems (MEMS) gyroscopes suffer from noise and sensitivity issues, and the actual data is usually corrupted by a varying bias and zero mean noise. The bias makes the expected mean of the gyroscope readings differ from the actual angular velocities. Although usually small in magnitude, the presence of bias in the data can corrupt odometry estimates, especially when the updates are in dead reckoning fashion. The small amount of bias builds up eventually and makes the odometry estimate deviate more and more from the actual value as time progresses. Thus, it is important to remove the inherent bias present in the angular velocity data before using it for linear velocity measurement purposes.

According to \cite{lakshminarayankalman}, the stochastic model of the angular rate measurement can be written as:

\begin{equation}\label{measurement model}
    G_{meas, k} = \omega + b + \epsilon_{k}
\end{equation}

where \(G_{meas, K}\) denotes the angular rate measurement of the gyroscope, \(\omega\) denotes the true gyroscope measurement, \(b\) is the bias drift and \(\epsilon_{k}\) represents the zero mean gyroscope noise. As shown in \cite{lakshminarayankalman}, we employ a Kalman filter to estimate the bias in the gyroscope measurements, with a minor deviation from the standard filter being a recursive approach for updating the observation model \(H\) and the pseudo measurement covariance \(R\) at each time step. The equations proposed are :

\begin{equation}\label{obs_model}
    H_{k} = 2(G_{meas, k}^{T} - b^{T})
\end{equation}
\begin{align}\label{pseudo_meas_cov}
    R_k &= 4\left(G_{meas,k} -b\right)^{T} \times \Sigma \times (G_{meas,k} - b) + 2 \times (tr\Sigma^{2})
\end{align}
    
Here, $G_{meas,k}$ is the gyroscope data measured at the $k_{th}$ time step and $b$ is the bias estimate updated at the previous time step. $\Sigma$ denotes the covariance matrix.

The rest of the measurement update equations are the same as that in  a standard Kalman filter as can be seen in \cite{lakshminarayankalman}.

\begin{figure}[!h]
\centering    
  \includegraphics[width=\linewidth]{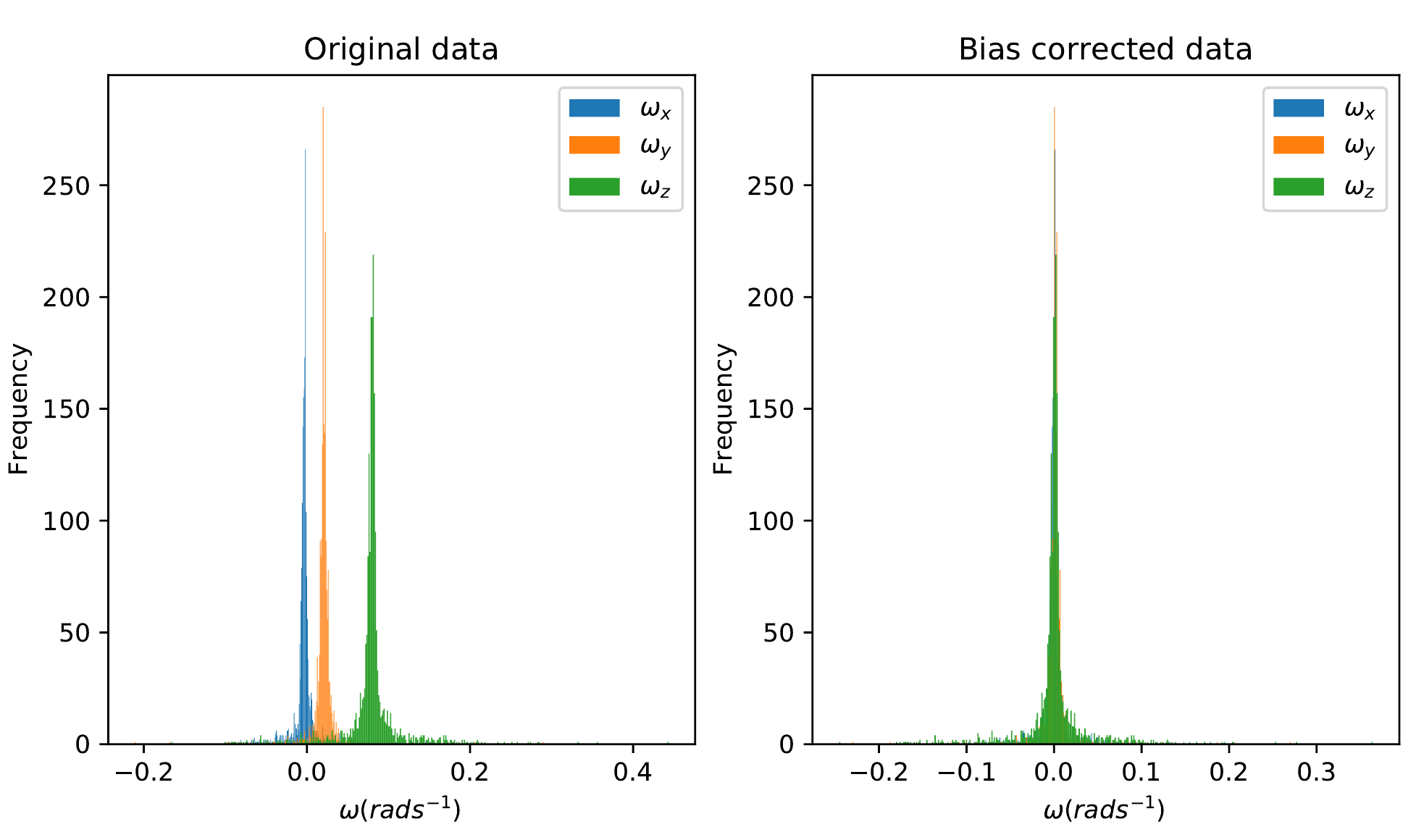}
  \caption{Removal of bias from data.}
  \label{fig:force}
  \label{biasremoval}
\end{figure}

Figure \ref{biasremoval} shows the result of the bias estimation and removal from the raw data. The gyroscope data was read in the absence of any motion. A Histogram of data was constructed with a precision of 1e-5 rad/sec, before and after the calculated bias was removed.

% Define abbreviations and acronyms the first time they are used in the text, 
% even after they have been defined in the abstract. Abbreviations such as 
% IEEE, SI, MKS, CGS, ac, dc, and rms do not have to be defined. Do not use 
% abbreviations in the title or heads unless they are unavoidable.

\subsection{Noise Reduction and Tracking True Angular Velocity}
The tracking of angular estimation is done using two approaches namely the Kalman filter and the RLS filter. 
\subsubsection{Kalman Filter}
Kalman Filter (KF) is further used to reduce the noise from the gyroscope data and estimate the true value of angular velocity. According to  \cite{xue2015noise}, $\dot b(t) = w_b(t)$, where $w_b(t)$ is noise of rate random walk (RRW) is also included in the stochastic model of angular rate measurement (\ref{measurement model}). True angular rate $\omega$ and the bias drift $b$ both are included in the state which would be estimated, and hence the KF state vector can be defined as X(t) = \([\omega , b]^{T}\). The KF state and measurement equation can be expressed as:

\begin{equation} \label{state measurement equation}
  \begin{aligned}
   \dot {\hat {X}} (t)= F \cdot X(t) + w(t)   \\
   Z(t) = H \cdot X(t) + v(t)
  \end{aligned}
\end{equation}

where \(H\) is the measurement matrix, \(F\) is the KF coefficient matrix,  \(Z(t)\) is the KF measurement information i.e. the output of gyroscopes, $w(t) = [n_w ,  w_b]^T$ is zero mean white noise vector, $v(t)$ is angular random walk(ARW) white noise.The measurement matrix $H$, covariance matrices $Q$ and $R$ can be given as  (\ref{matrices}):

\begin{equation}\label{matrices}
  H =\left[
  \begin{array}{cccc}
  1 & 1 \\
  \end{array}
\right], Q = \left[
  \begin{array}{cccc}
  q_w & 0 \\
  0 & q_b  \\
  \end{array}
\right] , R = q_n
\end{equation}

where where $q_b$ is the variance of RRW noise, $q_n$ is the variance of ARW noise, and $q_w$ is the variance of white noise $n_w$.The KF system is not completely observable as the rank of system observability matrix for the KF state-space model of equation (\ref{state measurement equation}) is equal to one. For off-line analysis of properties and characteristics of estimated covariance \(P(t)\) and filter gain \(K(t)\), a basic discrete iterative KF method \cite{xue2015noise} is used.

\begin{equation}\label{KF obs 1}
    P_{k|k-1} = F_{k|k-1}P_{k-1}F_{k|k-1}^{T} + Q_{k-1} \end{equation}
\begin{equation}\label{KF obs 2}
    K_{k} = P_{k|k-1}H_{k}^{T} \left(H_{k}P_{k|k-1}H_{k}^{T} + R_{k} \right)^{-1}
\end{equation}
\begin{equation}\label{KF obs 3}
    P_{k} = \left(1 - K_{k}H_{k} \right)P_{k|k-1}  \left(1 - K_{k}H_{k} \right)^{T} +  K_{k}R_{k}K_{k}^{T}
\end{equation}

The filter gain $K(t)$ approached a steady state value while $P(t)$ diverged without steady state values. The steady-state filter gain Ks obtained by the off-line approach of Equations (\ref{KF obs 1} - \ref{KF obs 3}) is defined as $K_s = [k_1 ,  k_2]^T$. Using this, the estimation of true angular rate can be obtained by a continuous-time KF from the following equation:

\begin{equation}\label{Kalman state update}
  \dot {\hat {X}} (t) = - \left[
  \begin{array}{cccc}
  k_1 & k_1 \\
  k_2 & k_2 \\
  \end{array}
\right]\hat X(t) +  \left[
  \begin{array}{cccc}
  k_1  \\
  k_2  \\
  \end{array}
\right] Z(t)
\end{equation}

Discretizing (\ref{Kalman state update}), and defining a matrix $m=\left[
  \begin{array}{cccc}
  k_1  \\
  k_2  \\
  \end{array}
\right]$ , then we have:

\begin{equation}\label{Kalman state update- discrete}
    \hat X_{k+1} = e^{-mT}\hat X_{k} + \int_{0}^{T} e^{-mt} dt  \left[
  \begin{array}{cccc}
  k_1  \\
  k_2  \\
  \end{array}
\right] Z(t)
\end{equation}

From the eigenvalue decomposition of matrix $m = S\Lambda S^{-1}$, where S is a matrix with columns corresponding to eigenvectors and $\Lambda$ is a diagonal matrix of eigenvalues. From the definition of matrox $m$, the two eigenvalues of m calculated are $\lambda_1 = \left(k1 + k2\right)$ and $\lambda_2 = 0$, then we get:
\begin{equation} \label{integral value}
\begin{split}
\int_{0}^{T} e^{-mt} dt &= \int_{0}^{T} e^{-S\Lambda S^{-1}t} dt  \\ & = S  \left[
  \begin{array}{cccc}
  -\lambda_1^{-1}(e^{-\lambda_1 T } - 1) & 0 \\
  0 & T 
  \end{array}
\right] S^{-1}   
\end{split}
\end{equation}

where \(T\) is the sampling time. From the equations (\ref{Kalman state update- discrete}) and (\ref{integral value}), the true angular velocity and the bias drift can be estimated as (\ref{final estimation})

\begin{equation} \label{final estimation}
  \begin{aligned}
   \hat \omega_{k+1} = \left[
  \begin{array}{cccc}
  1 & 0 \\
  \end{array}
\right] \cdot \hat X_{k+1}
   \\
   \hat b_{k+1} = \left[
  \begin{array}{cccc}
  0 & 1\\
 \end{array}
\right] \cdot \hat X_{k+1}
  \end{aligned}
\end{equation}

% \begin{equation}
% \hat{X}(t) = \left(\begin{array}{cc} k_1 & k_1\\ k_2 & k_2 \end{array}\right)

% \end{equation}

\subsubsection{Recursive Least Squares Filter}
Recursive least squares (RLS) \cite{haykin2008adaptive} is a popular adaptive filter algorithm that finds the cost-minimizing coefficients recursively. The cost function is the weighted linear least squares related to the input data. The performance in terms of convergence rate, tracking, misadjustment, and stability depends on the forgetting factor $\left(0<\lambda<1\right)$. The forgetting factor is used to discount the past data and hence acts as a bias-variance trade-off parameter. After initializing the bias as the estimated static bias, the model parameter is updated on arrival of every measurement of the angular velocity $\omega_z$. The model-predicted value of $\omega_z$ is corrected using the bias which is estimated in the same iteration by incorporating it into the state vector\cite{haykin2008adaptive}.

\begin{equation}\label{RLS_Update}
    \hat \beta_{n+1} = \hat \beta_n + P_{n+1}{x_{n+1}}^\top\left(y_{n+1}-x_{n+1}\hat \beta_n\right)
\end{equation}
\begin{equation}\label{RLS_cov_update}
    P_{n+1} = \frac{1}{\lambda}\left(P_n - \frac{P_n {x_{n+1}}^\top x_{n+1} P_n}{\lambda + x_{n+1} P_n {x_{n+1}}^\top}\right)
\end{equation}

The equations (\ref{RLS_Update}) and (\ref{RLS_cov_update}) govern the state estimation using RLS where $y_n = \omega_{z_n}$ and $ x_n = [n,\hat y_{n-1} - y_n]^\top $.

\begin{figure*}[h]
\centering
\noindent \includegraphics[width=0.7\linewidth]{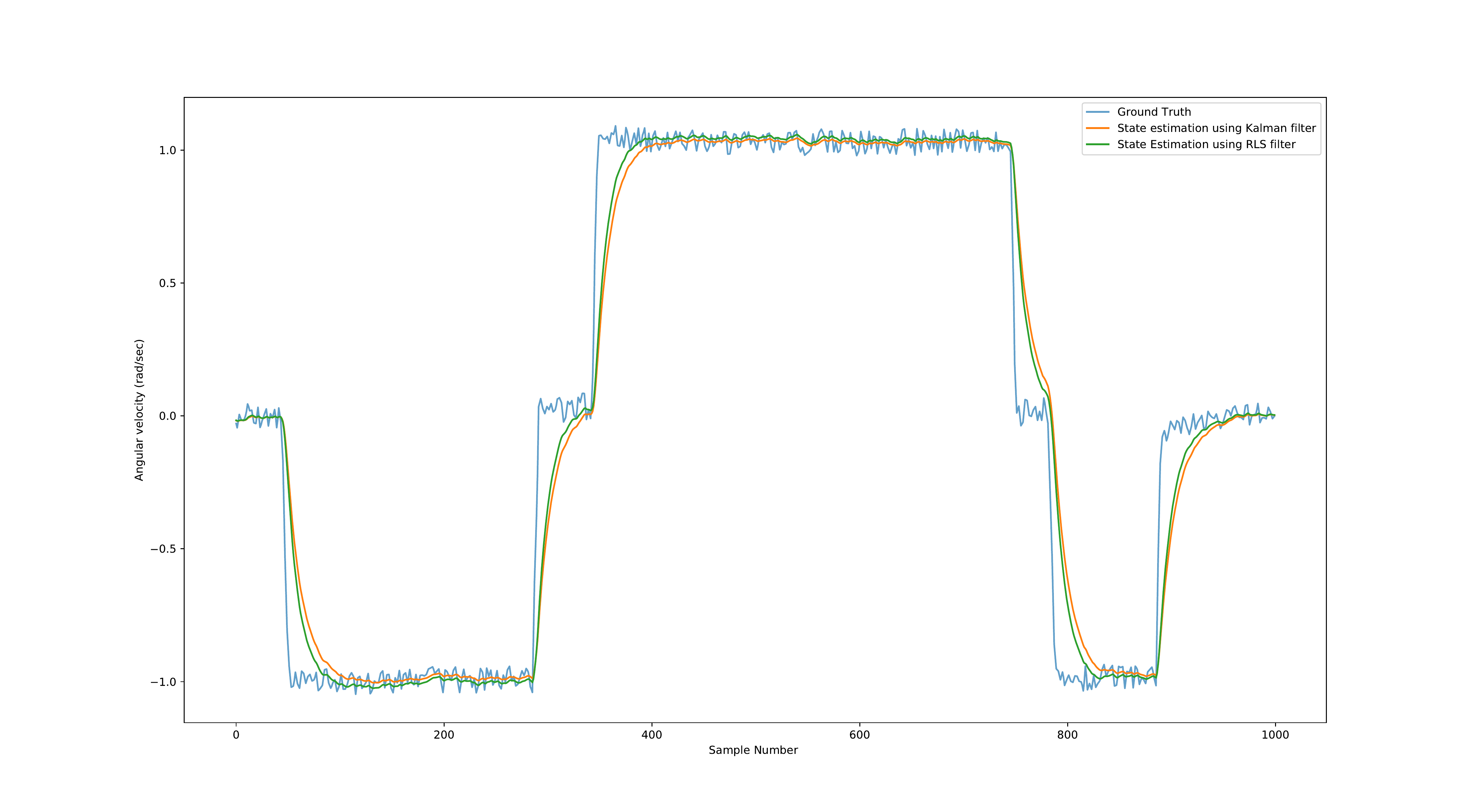}
\caption{Evaluation of performance of Kalman Filter and RLS on simulation data}
\label{comparison}
\end{figure*}

Figure \ref{comparison} compares the performance of state estimation using Kalman filter and Recursive Least Square methods. The data is generated synthetically using a robot simulator implemented in ROS and Gazebo, to compare results of bias estimation with actual bias inserted into the system.
% [Values to be added.]

% \subsection {Using the Gyroscope data}

% After  the  bias  estimation  and  noise  reduction  the  components of the angular velocity obtained from gyroscope are used to measure the  linear and the angular velocity of the robot. Assuming the plane of the the gyroscope to be parallel to the plane  of  the  wheel  and  perpendicular to the plane of ground the z component of the gyroscope gives the rpm of the wheel. Using data gx, gy - IMU. The x and y components of the gyroscope are used to find the angular velocity of the robot using the following relation

\section {Factor graph-based Localization}
\label{sec:localize}
We propose a factor graph based localization scheme for the ground vehicle. We choose to constrict our ground vehicle's state to be defined by only three variables, i.e. [x,y and \(\theta\)]. This approximation drastically reduces the complexity of the localization, while not impacting the estimated quality by much.

A factor graph is a probabilistic graphical model, in which a graph of unknown and known variables (or states) is constructed, connected by factors which encode probabilistic information in them. These factors combine the known and unknown variables and define the relationship between them. Maximum a-posteriori (MAP) estimate can then be generated by maximizing the factor product \cite{loeliger2004introduction}.

\begin{equation} \label{factor}
f(X_1, X_2, \dots X_n) = \prod{f_i(\chi_i)}
\end{equation}
where \(X_1, X_2, \ldots X_n\) represent the variables in the graph,  \(f_i\) are the factors and \(\chi_i\) are variables connected to the \(i^{th}\) factor. 

We use data from the two wheel odometers, GPS data and compass data to construct the factor graph. Such graphs can be modeled very easily in the Georgia Tech Smoothing And Mapping (GTSAM) C++ library. For our purposes, we need just two kinds of factors, a 3 degree-of-freedom odometry factor that connects successive odometry variables and a 2D GPS factor, which calculates the jacobian between a SE(2) Pose2 and angleless (X,Y) coordinates. The GPS measurements (~5Hz) are much slower than the odometer measurements (~100Hz), hence there are many different odometry readings between two consecutive GPS readings. Since GPS readings are encoded as latitude and longitude which are in degrees, we need to first convert them into a local coordinate system to encode the motion in meters. For this, we employ the use of Universal Transverse Mercator (UTM) coordinate system \cite{wikiutm}, which divides the earth into 60 total divisions, and returns the data in the form of a northing and easting in meters. This data, although approximate, can be used to encode robot motions in real world coordinates using Latitude and Longitude data input. The relative odometry is calculated using dead reckoning as given by equation (\ref{deadreckoning}) \cite{siegwart2011introduction}.

\begin{equation}\label{deadreckoning}
  \begin{aligned}
  &\Delta s = \frac{\Delta s_{l} + \Delta s_{r}}{2} \\
  &\Delta \theta = \frac{\Delta s_{r} - \Delta s_{l}}{b} \\
  &\left[ {\begin{array}{ccc}
   x_{n+1}  \\
   y_{n+1}  \\
   \theta_{n+1} \\
  \end{array} } \right] = \left[ {\begin{array}{ccc}
   x_{n}  \\
   y_{n}  \\
   \theta_{n} \\
  \end{array} } \right] + \left[ {\begin{array}{ccc}
   \Delta s \times cos(\theta_{n} + \frac{\Delta \theta}{2})  \\
   \Delta s \times sin(\theta_{n} + \frac{\Delta \theta}{2})  \\
   \Delta \theta \\
  \end{array} } \right]
  \end{aligned}
\end{equation}

%http://web.eecs.utk.edu/~leparker/Courses/CS594-fall08/Lectures/Nov-13-Localization-Mapping-I.pdf

\begin{figure}[h] 
\centering
\noindent \includegraphics[width=0.9\linewidth]{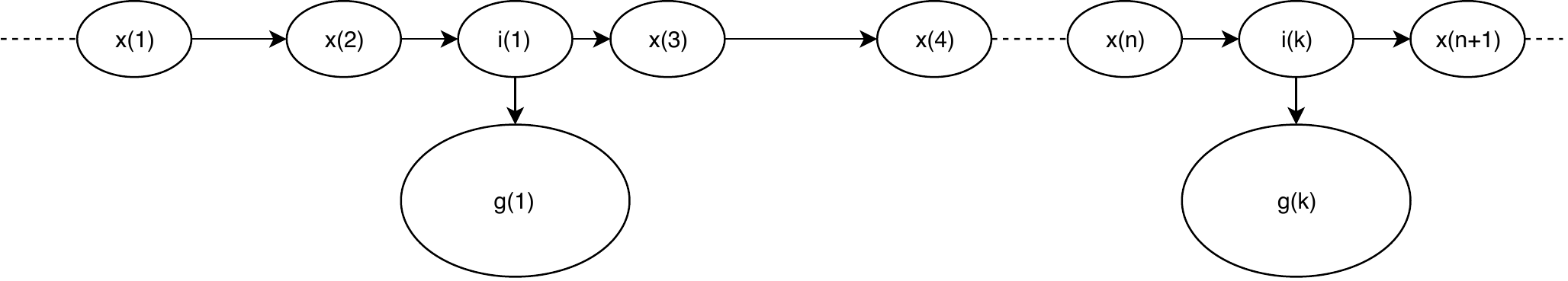}
\caption{Example factor Graph for localization}
\label{factorgraphloc}
\end{figure}

where \(\theta\) is the magnetic yaw angle given by an INS (Inertial Navigation System) unit placed on the vehicle.

As shown in Figure \ref{factorgraphloc}, when the \(k^{th}\) GPS data arrives, the relative pose factor between the previous \(x_{n}\) and next state \(x_{n+1}\) is linearly interpolated using the timestamps of the GPS message and a new state \(i_{k}\) is added at the timestamp of the GPS data. A large factor graph is thus constructed. The horizontal arrows are all relative odometry \textit{BetweenFactor}s and the vertical arrows are \textit{GPSPose2Factor}s. Each relative odometry factor is given a variance of 10\% of the value and each GPS factor is given a fixed variance of 5m in both northing and easting. The graph is solved using a GaussNewton solver to find the maximum a-posteriori (MAP) estimate \cite{rosen2012incremental}.

\begin{figure*}[h]
\centering
\noindent \includegraphics[width=0.6\linewidth]{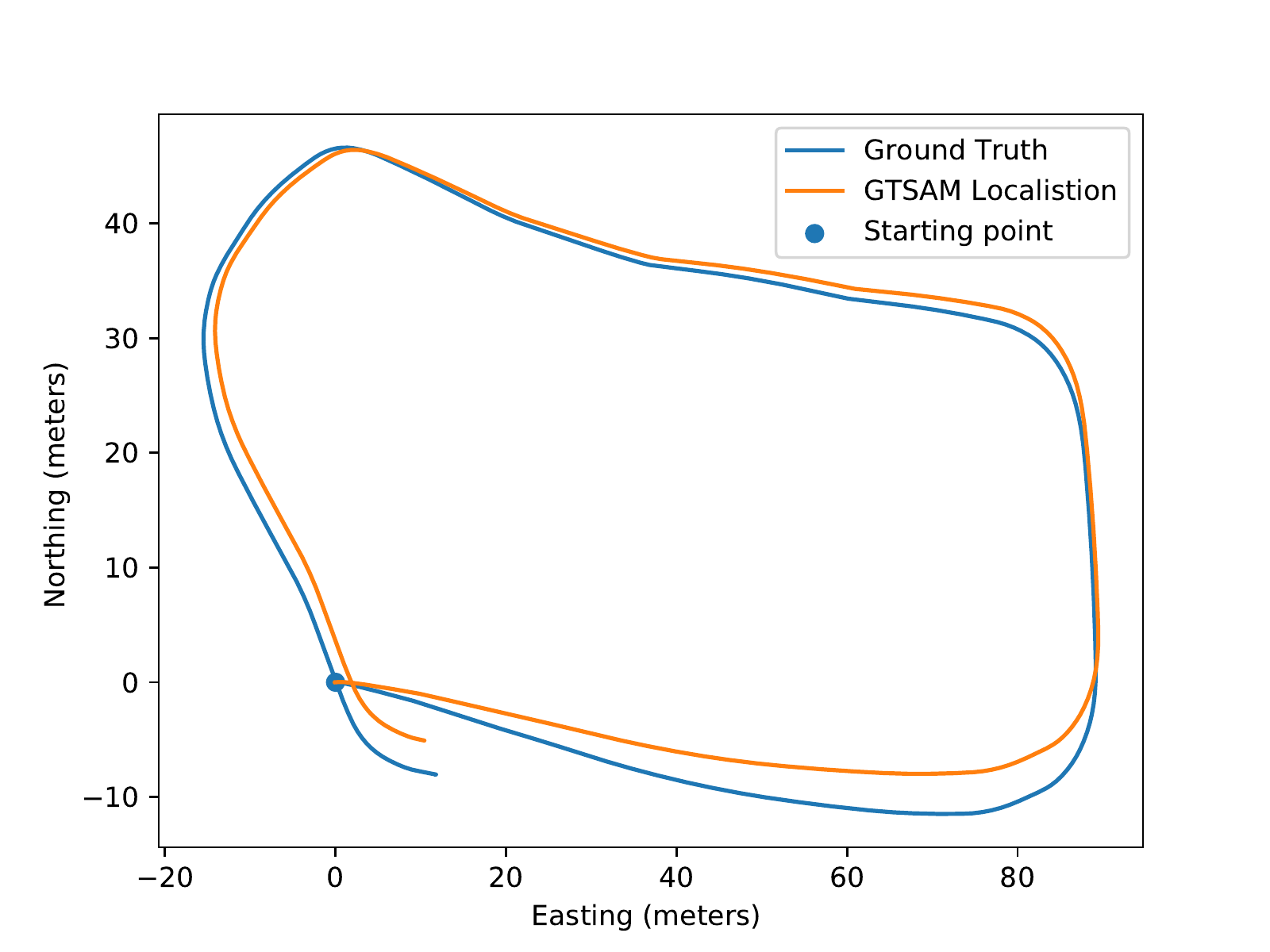}
\caption{Comparison of Factor graph based odometry result with Husky's odometry package}
\label{localization}
\end{figure*}

\section {Experiments and Results}
\label{sec:experiment}
\begin{figure*}[h] 
\centering
\noindent \includegraphics[width=0.9\linewidth]{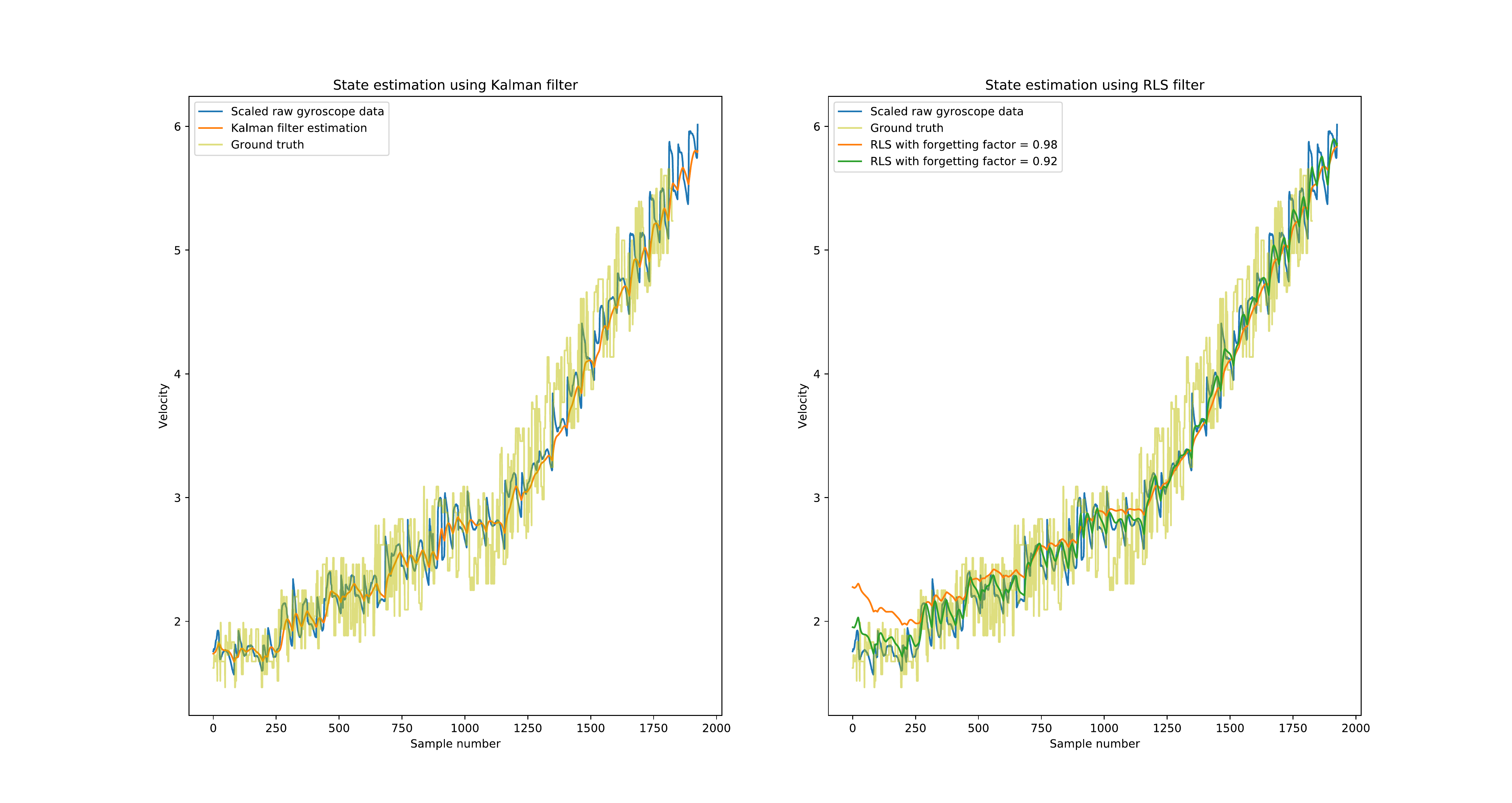}
\caption{Comparison of Kalman Filter and RLS based state estimation on real data (Unit of the velocity recorded is kmph) }
\label{comparisongt}
\end{figure*}
As described in section \ref{sec:nonintrusive}, the non-intrusive inertial speed sensor has been aggregated using Inertial Measurement Units, wireless  communication modules, microcontrollers, batteries,  power converters and charging modules.

\begin{figure}[t] 
\begin{center}
\includegraphics[width=0.7\linewidth]{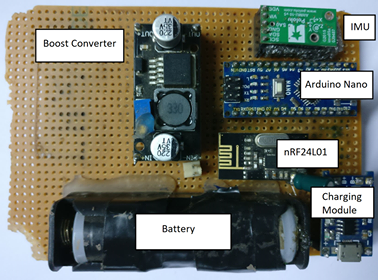} 
\includegraphics[width=0.7\linewidth]{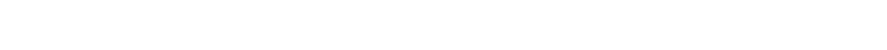}
\includegraphics[width=0.7\linewidth]{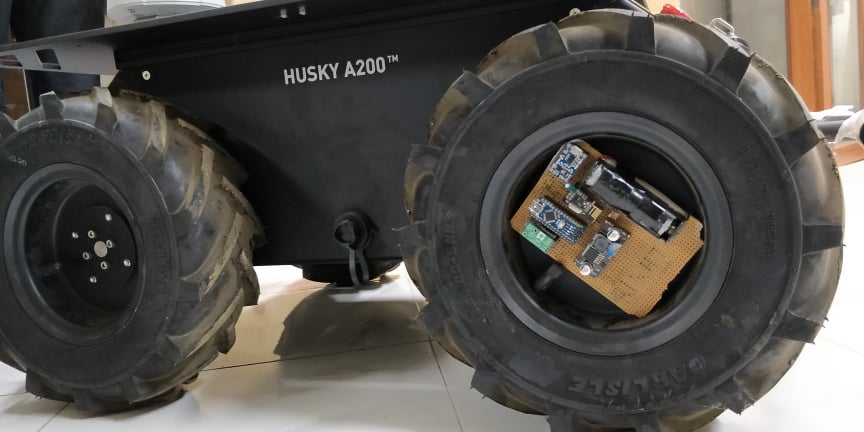}
\caption{Top: Prototype of the speed sensing module, Bottom: The module attached to the rear wheel of Husky robot}
\label{hardware}
\end{center}
\end{figure}

We have used the Pololu MinIMU-9 as an inertial sensor. It has a 3 axis gyroscope with an I\(^{2}\)C interface for communication. It is a low-cost MEMS IMU that outputs gyroscope readings at a sensitivity of upto \(\pm2000^o\)/sec, at rates of upto 1.6KHz. It is also incredibly power efficient, the total power consumption of the IMU sits below 10mW even under load. Each IMU is interfaced to an Arduino Nano microcontroller, which is preferred to larger boards like Arduino Uno since we want to limit the circuit to as small a size as possible. We have used a nRF24L01 module for wireless transmission of data from the transmitter to the receiver. The receiver setup consists of another Arduino with a nRF24L01 module attached to it, hooked to the processing computer. The USB-serial data transmission is handled by ROSSerial, which publishes the speeds of both left and right wheels over ROS topics. Since the entire setup needs to be wireless, it is provided with a 18650-sized 3.3V Lithium Ion battery. A charging circuit in the form of TE585 lithium battery charger module is also placed on the board. The hardware is shown in Figure \ref{hardware}. The GPS and compass yaw readings are provided by a VectorNav VN-100 Rugged sensor, which is mounted on the body of the vehicle. It is an Inertial Measurement Unit(IMU) and Attitude Heading Reference System(AHRS) that combines 3-axis accelerometers, 3-axis gyros, 3-axis magnetometers. a barometric pressure sensor and a 32-bit processor.It is a high-performance system with \(\pm16\)g accelerometer range and \(\pm2000^o\)/sec gyroscope range with data output rates of upto 800Hz. The GPS accuracy in India is usually lower than in countries like USA, and GPS data is usually off by at least 5m, even when triangulation is done using 8 satellites. Thus, the importance of Factor graph-based localization is very high. 

For testing, 2 non-intrusive speed sensors were mounted on the rear wheels of a Husky, an industrial grade robot made by Clearpath Robotics. Husky has high-grade optical encoders that we consider as ground truth. Husky was moved (using teleoperation) around a loop of length ~250m and the performance of the RLS and Kalman filters was recorded, in comparison with actual wheel encoders. The resultant graph is shown in Figure \ref{comparisongt}. As evident from the graph, both Kalman and RLS filters are able to track the gyroscope data and ground truth (optical encoder readings) satisfactorily. In fact, we observe much lower white noise content than both raw gyroscope data and optical encoder data. Then, the data from RLS was used for dead reckoning and a plot of the trajectory traced by the robot was generated. As evident from Figure \ref{localization}, the error in localization is close to 1.5 percent of the total path length. Please note that installation of actual ground truth measurement devices like motion capture systems was not possible for us, and hence we simply consider the Husky odometry output to be ground truth, which might explain the relatively higher deviation between the two trajectories in Figure \ref{localization}.

\section{Conclusion}
\label{sec:conclusion}
This paper demonstrated a non-intrusive sensor which is reliable enough to replace a regular wheel encoder in a ground vehicle. It was shown that this sensor can be accurately used to localize the robot as well. This current implementation is mounted on the Mahindra E2O electric vehicle to allow cruise control and trajectory tracking algorithms to function, and the performance is qualitatively much better than algorithms using the default speedometer (1kmph resolution). In near future, we plan to develop a System-on-Chip(SoC) for integrating the embedded components, for making the overall module more compact and easier to attach on any vehicle or robot.
\section*{Acknowledgement}

We would like to thank team Autonomous Ground Vehicle, IIT Kharagpur and Sponsored Research \& Industrial Consultancy, IIT Kharagpur for providing us with resources and funding for this project. We would also like to thank Mahindra Automobiles for providing the test vehicle to us. We would also like to acknowledge the efforts put in by Gopabandhu Hota, Dishank Yadav, Shubham Sahoo and Rishabh Singh during this task.

\bibliography{research}{}
\bibliographystyle{IEEEtran}
 
\vspace{12pt}
\color{red}
% IEEE conference templates contain guidance text for composing and formatting conference papers. Please ensure that all template text is removed from your conference paper prior to submission to the conference. Failure to remove the template text from your paper may result in your paper not being published.

\end{document}